\documentclass[preprint,nofootinbib,showpacs,12pt]{revtex4-1}

\usepackage{amssymb}
\usepackage{graphicx}
\usepackage{dcolumn}
\usepackage{bm}
\usepackage{color}
\usepackage{latexsym}
\usepackage{lineno}
\usepackage{comment}
\usepackage{color}

\begin{document}
\newcommand{\leftg}{\langle \phi_0 |}
\newcommand{\rightg}{| \phi_0 \rangle}
\newcommand{\chiral}{\langle \bar{q} q \rangle} 
\newcommand{\vs}{\vspace{-0.25cm}}
\newcommand{\gev}{\,\mathrm{GeV}}
\newcommand{\mev}{\,\mathrm{MeV}}
\newcommand{\fmd}{\,\mathrm{fm}^{-3}}
\newcommand{\ord}[1]{\mathcal{O}(k_f^{#1})}
\newcommand{\dif}{\mathrm{d}}

\title{Nuclear Pairing from Chiral Pion-Nucleon Dynamics:
Applications to Finite Nuclei
}

\author{Paolo Finelli}
\affiliation{Dipartimento di Fisica, Universit\`a di Bologna, I-40126 Bologna, Italy\\
INFN, Sezione di Bologna, I-40126 Bologna, Italy}
\author{Tamara Nik{\v s}i{\' c}}
\author{Dario Vretenar}
\affiliation{Physics Department, Faculty of Science, University of
Zagreb, HR-10000 Zagreb, Croatia}

\begin{abstract}
The $^1S_0$ pairing gap in isospin-symmetric nuclear matter and finite nuclei 
is investigated using the  chiral nucleon-nucleon potential at the N$^3$LO order 
in the two-body sector, and the N$^2$LO order in the three-body sector.
To include realistic nuclear forces in RHB (Relativistic Hartree Bolgoliubov) 
calculations we rely on a 
separable form of the pairing interaction adjusted to the bare nuclear force.
The separable pairing force is applied to the analysis 
of pairing properties for several isotopic and isotonic chains
of spherical nuclei.
\end{abstract}

\maketitle

\section{Introduction}

Nuclear energy density functionals (EDF) provide a microscopic, 
globally accurate, and yet economic description of ground-state properties 
and collective excitations over the whole nuclide chart. 
Even though it originates in the effective interaction between nucleons, a 
generic density functional is not necessarily related 
to any given nucleon-nucleon (NN) potential and, in fact, some of the most 
successful modern functionals are entirely empirical \cite{Bender:2003jk}.
Of course it is very much desirable to have a fully microscopic foundation for a 
universal density functional, and this is certainly one of the major 
challenges in low-energy nuclear structure physics \cite{review}. 

The EDF approach to nuclear structure is analogous to Kohn-Sham density 
functional theory (DFT) and, because it includes correlations, goes beyond the 
Hartree-Fock approximation. Kohn-Sham DFT has the advantage of being a local 
scheme, but its usefulness crucially depends on our ability to 
construct accurate approximations for the most important part of the 
functional, that is, the universal exchange-correlation functional \cite{DG.90}. 

In a series of recent articles \cite{FKV.03,FKV.04,FKV.06} concepts of
effective field theory and density functional theory have been used to derive
a microscopic relativistic EDF-based model of nuclear many-body dynamics
constrained by in-medium QCD sum rules and chiral symmetry. The
density dependence of the effective nucleon-nucleon couplings in this
model (called FKVW in the following) is determined from the long- and intermediate-range
interactions generated by one- and two-pion exchange processes. 
They are computed using in-medium chiral
perturbation theory, explicitly including $\Delta(1232)$ degrees of freedom
\cite{Fri.04}. 
Divergent contributions to the nuclear matter energy density, calculated 
at three-loop level, are absorbed by few contact terms. These constants  
are understood to encode unresolved short-distance dynamics. 

The relativistic FKVW model has successfully been employed in studies 
of ground-state properties of spherical and deformed nuclei using the 
Relativistic Hartree-Bogoliubov framework (RHB \cite{VALR.05}). In the 
description of open-shell nuclei, in particular, a hybrid model has been used 
with the FKVW Kohn-Sham potential in the particle-hole ($ph$) channel and, like in most 
applications of RHB-based models,  the pairing part of the empirical Gogny 
force~\cite{BGG.91} in the particle-particle ($pp$) channel. 

Even though this approach has been very successful, it is not theoretically 
consistent because of the choice of the empirical effective interaction in the 
$pp$ channel. Quite recently, as part of a larger program to develop a framework of  
fully microscopic nuclear energy density functionals, much effort has  
been devoted to designing non-empirical pairing 
functionals \cite{Duguet:2007be,Les.09,Heb.09, Pair3b,Lesinski:2011rn}. 

The aim of this work is to formulate a consistent microscopic framework for open-shell nuclei, 
in which both the {\it ph} and the {\it pp} channels of the effective inter-nucleon 
interaction are determined by chiral pion-nucleon dynamics. Thus we 
consider a separable $pp$ interaction based on a microscopic pairing 
interaction constrained by chiral dynamics (see Ref. \cite{KNV.05} for  
previous calculations involving the N$^2$LO chiral potential), combine it with the 
FKVW functional in the $ph$ channel and, employing the 
corresponding RHB model, present a study of pairing gaps in isotopic and isotonic 
chains of spherical open-shell nuclei. 

We will use the realistic NN potential developed by the Idaho group
at next-to-next-to-next-to-leading order (N$^3$LO) in the chiral expansion \cite{Machleidt:2011zz}
(see also Ref. \cite{Epelbaum:2008ga}),
and a two-body density-dependent potential derived from the relevant
diagrams at the N$^2$LO order in the three body sector \cite{Holt:2009ty} 
(see also Refs. \cite{Hebeler:2009iv,Hebeler:2010xb,Lesinski:2011rn,Gandolfi,Lovato:2010ef} for similar approaches and 
pertinent details).

The paper is organized as follows. In Sec. \ref{sec_inf} we discuss results for the pairing gap of nuclear matter 
in the BCS approximation. Sec. \ref{sec_tyan} briefly reviews the method introduced by Y. Tian {\it et al.} 
\cite{TMR.09a, TMR.09b} to apply realistic pairing interactions to calculations of finite nuclei. In Sec. \ref{sec_res} we 
analyze pairing gaps in spherical nuclei for several isotopic and isotonic chains. Sec. \ref{sec_con} summarizes the 
principal results.

\section{Pairing gap in a homogeneous infinite system}
\label{sec_inf}

The momentum and density-dependent pairing field 
$\Delta(k,k_F)$ in infinite matter is determined by the solution of 
the BCS gap equation
\begin{equation}
\label{gapeq}
\Delta(k,k_F) = -\frac{1}{4\pi^2}
  \int_0^\infty{\frac{p^2 V(p,k) \Delta(p,k_F)}
   {\sqrt{[ {\cal E}(p,k_F)-{\cal E}(k_F,k_F)]^2+\Delta(p,k_F)^2}}\; {\rm d}p} \;,
\end{equation}     
where $V(p,k)$ represents the off-shell pairing potential in momentum space,
${\cal E}(p,k_F)$ is the quasiparticle energy, and ${\cal E}(k_F,k_F)$ is the
Fermi energy. 

The effective force in the pairing channel is, in principle, generated by the sum
of all particle-particle irreducible Feynman diagrams~\cite{Mig.67}. 
In most application to nuclear and neutron matter, however, 
only the lowest-order term, that corresponds to the bare nucleon-nucleon 
interaction, is retained~\cite{DH-J.03}. Terms of higher order in the effective
pairing interaction represent screening corrections to the bare force, caused by
medium polarization effects (see Refs. \cite{Sch.96,Lombardo:2005sw} and references therein). 
In the present analysis we only consider the bare interaction, 
while a study of polarization effects will be carried out in a forthcoming paper.

For the pairing potential  $V(p,k)$ we employ the simple ansatz:
\begin{equation}
\label{V_pk}
V(p,k) = V_{2B} (p,k) + V_{3B}(p,k,m) \simeq V_{2B} (p,k) + \bar{V}_{2B} (k_F,p,k) \; ,
\end{equation}
where the three-body potential is approximated by an effective two-body density-dependent
potential $\bar{V}_{2B}$ derived by Holt {\it et al.} in Ref \cite{Holt:2009ty}. 
These authors showed 
that in the singlet channel ($^1S_0$) the overall effect of $\bar{V}_{2B} (k_F,p,k)$ 
is to reduce the strong S-wave attraction (cf. Fig. 6 of Ref. \cite{Holt:2009ty}). 
As suggested in Ref. \cite{Holt:2009ty}, here we neglect a possible isotopic dependence
that, in any case, is expected to be rather small for the nuclei considered in the present analysis 
(see also Ref. \cite{Lesinski:2011rn}).
For both terms in Eq.~(\ref{V_pk}) we follow standard procedures for the regulator functions, and refer
the reader to the original papers for details.

For the single-particle spectrum that appears in the denominator of the gap equation (\ref{gapeq}) 
we employ the simple quadratic form
\begin{equation}
{\cal{E}} (p, k_F) - {\cal{E}} (k_F,k_F) = \frac{p^2 -k_F^2}{2M^\ast(k_F)} \;.
\end{equation}
This approximation should suffice because the momenta $p$ around $k_F$ give the dominant contribution to the 
integral in Eq. (\ref{gapeq}). The effective nucleon mass $M^\ast(k_F)$ was obtained in 
a very recent calculation (Fig. 9 of Ref. \cite{Holt:2011nj}), in which the nuclear energy density functional was 
derived to first order in the two- and three-nucleon interaction using a density matrix expansion\footnote{In Ref. \cite{Holt:2011nj}
the two-body interaction comprises long-range one- and two-pion exchange contributions and a set of contact terms contributing up to fourth power in momenta (N$^3$LOW potential developed
lowering the cut-off scale to $\Lambda= $ 414 MeV). 
In addition, the authors employ the leading order chiral three-nucleon interaction with 
the corresponding parameters $c_E$, $c_D$ and $c_{1,3,4}$ adjusted in calculations of few-body systems. Even though the results are in good agreement with previous calculations \cite{Fritsch:2004nx}, one should note that higher-order corrections could have non-negligible effects \cite{Holt:2011nj2}.
}.

\begin{figure}[h]
\begin{center}
\includegraphics[scale=0.6,angle=0.0]{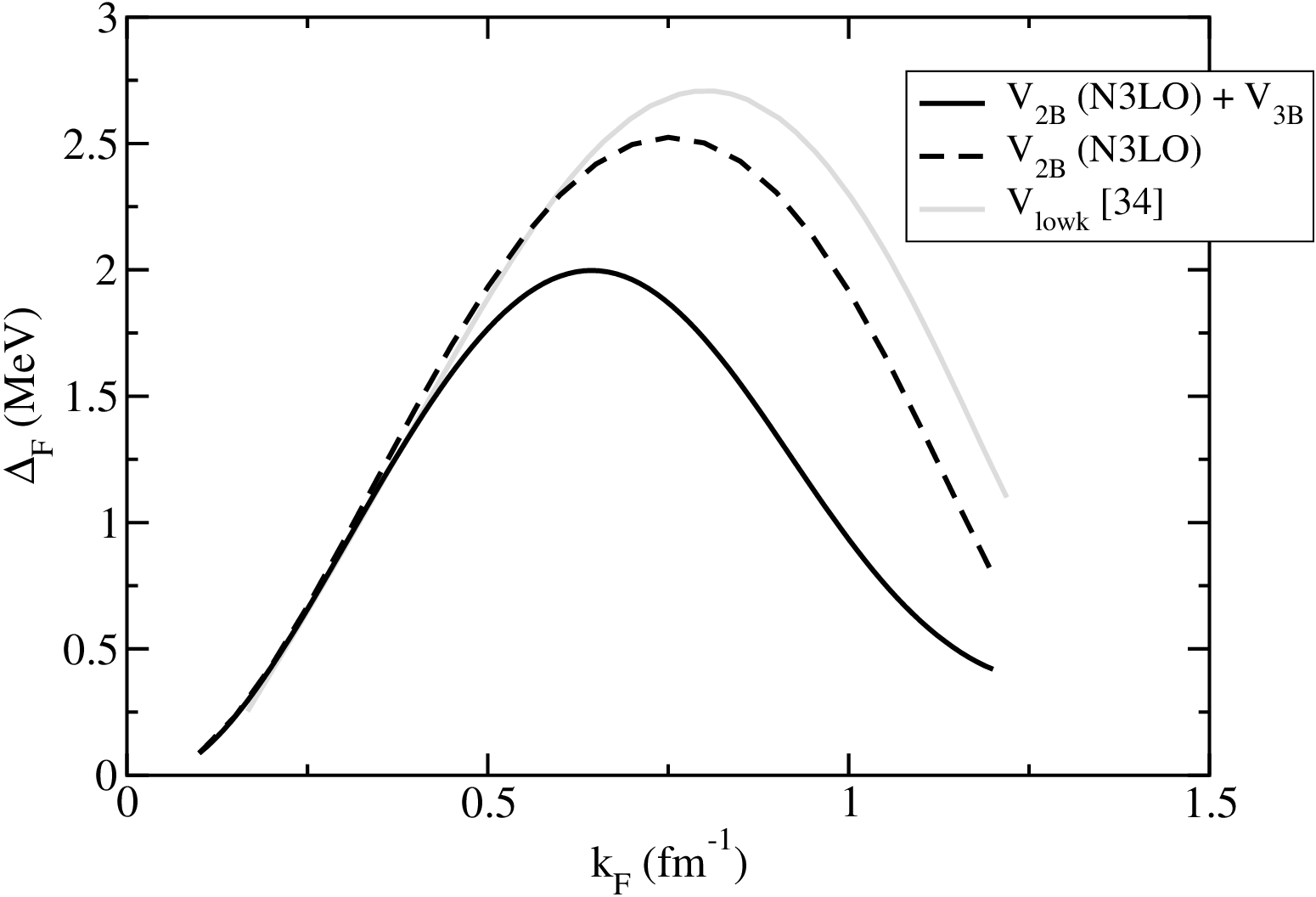}
\caption{\label{fig1}
Pairing gaps in symmetric nuclear matter in the $^1S_0$ channel as functions of
the Fermi momentum. The dashed curve is obtained by including only the 
two-body ($V_{2B}$) chiral interaction, whereas the solid curve  also includes
the contribution of  three-body ($V_{3B}$) forces. The gaps are compared to those 
obtained in Ref. \cite{Sedrakian:2003cc} using the 
$V_{\rm lowk}$ potential.}
\end{center}
\end{figure}

Fig. \ref{fig1} displays the pairing gap $\Delta(k_F,k_F)$ in symmetric nuclear matter as function 
of the Fermi momentum $k_F$. We plot results of the complete calculation that
includes two and three-body forces (solid curve), and the pairing gap obtained 
with only the two-body NN potential at N$^3$LO (dashed curve). Our results are 
shown in comparison with those obtained in Ref. \cite{Sedrakian:2003cc} using the 
$V_{\rm lowk}$ potential (with single
particle energies computed in Brueckner-Hartree-Fock theory).

\section{Mapping procedure}
\label{sec_tyan}
To implement the chiral NN potential at N$^3$LO in 
the pairing channel of the RHB framework for finite nuclei, we adopt the 
approach of Refs.~\cite{TMR.09a,TMR.09b} where a  
separable form of the pairing interaction was
introduced, with parameters adjusted to reproduce the pairing properties
of the Gogny force in nuclear matter. 
In nuclear matter the pairing force is separable in momentum 
space:
\begin{equation}
\left\langle {\bm k}\right\vert V^{^{1}S_{0}}\left\vert {\bm k}^{\prime}\right\rangle
=-Gp(k)p(k^{\prime})\;. \label{sep_pair}%
\end{equation}
By assuming a simple Gaussian ansatz $p(k)=e^{-a^{2}k^{2}}$, the two
parameters $G$ and $a$ have been adjusted to reproduce the density dependence
of the gap at the Fermi surface, calculated with a Gogny force. For the D1S
parameterization~\cite{BGG.91} of the Gogny force: $G=728\;\mathrm{MeVfm}%
^{3}$ and $a=0.644\;\mathrm{fm}$. Here we apply the same procedure to 
the chiral NN potential at the N$^3$LO order. 
For finite nuclei, when the pairing force 
Eq.~(\ref{sep_pair}) is transformed from momentum to coordinate space, it takes the form:
\begin{equation}
V({\mbox{\boldmath $r$}}_{1},{\mbox{\boldmath $r$}}_{2},{\mbox{\boldmath $r$}}%
_{1}^{\prime},{\mbox{\boldmath $r$}}_{2}^{\prime})=G\delta\left(
{\mbox{\boldmath $R$}}-{\mbox{\boldmath $R$}}^{\prime}\right)
P(r)P(r^{\prime})\frac{1}{2}\left(
1-P^{\sigma}\right)  , \label{pp-force}%
\end{equation}
where ${\mbox{\boldmath $R$}}=\frac{1}{2}\left(  {\mbox{\boldmath $r$}}%
_{1}+{\mbox{\boldmath $r$}}_{2}\right)  $ and ${\mbox{\boldmath $r$}}%
={\mbox{\boldmath $r$}}_{1}-{\mbox{\boldmath $r$}}_{2}$ denote the
center-of-mass and the relative coordinates, $P(r)$
is the Fourier transform of $p(k)$:
\begin{equation}
P(r)=\frac{1}{\left(  4\pi a^{2}\right)  ^{3/2}%
}e^{-r^{2}/4a^{2}}\;, \label{P3D}%
\end{equation}
and the factor $1/2\left(1-P^{\sigma}\right)$ projects on the $^{1}S_0$ channel.
The pairing force has a finite range and, because of the presence of the factor
$\delta\left(  {\mbox{\boldmath $R$}}-{\mbox{\boldmath
$R$}}^{\prime}\right)  $, it preserves translational invariance. Even though
$\delta\left(  {\mbox{\boldmath $R$}}-{\mbox{\boldmath
$R$}}^{\prime}\right)  $ implies that this force is not completely separable
in coordinate space, the corresponding anti-symmetrized $pp$ matrix elements
can be represented as a sum of a finite number of separable terms, using a 
method developed by Talmi and Moshinsky. When the nucleon wave functions 
are expanded in a harmonic oscillator basis, spherical
or deformed, the sum converges relatively quickly.  A relatively 
small number of separable terms reproduces with high accuracy the result of
a calculation performed in the complete basis. We refer 
the reader to \cite{TMR.09a} for more details. 

The parameters of the separable pairing force take the values:
$G = 892.0$ MeVfm$^{3}$ and $a= 0.74$ fm for the N$^3$LO potential, 
and $G =1045.0$ MeVfm$^{3}$ and $a= 0.86$ fm for the complete potential $V(p,k)$. We note that
recently a similar approach was employed in Refs. \cite{Les.09,Lesinski:2011rn}, where
a low-rank separable representation was used to reproduce directly 
$V_{\rm low k}$ and $V_{3N}$  in the $^1S_0$ channel 
(for $V_{3N}$ the density dependence was parametrized by a polynomial in the Fermi momentum).  

\section{Results for finite nuclei}
\label{sec_res}

\begin{figure}[h]
\begin{center}
\includegraphics[scale=0.6,angle=0.0]{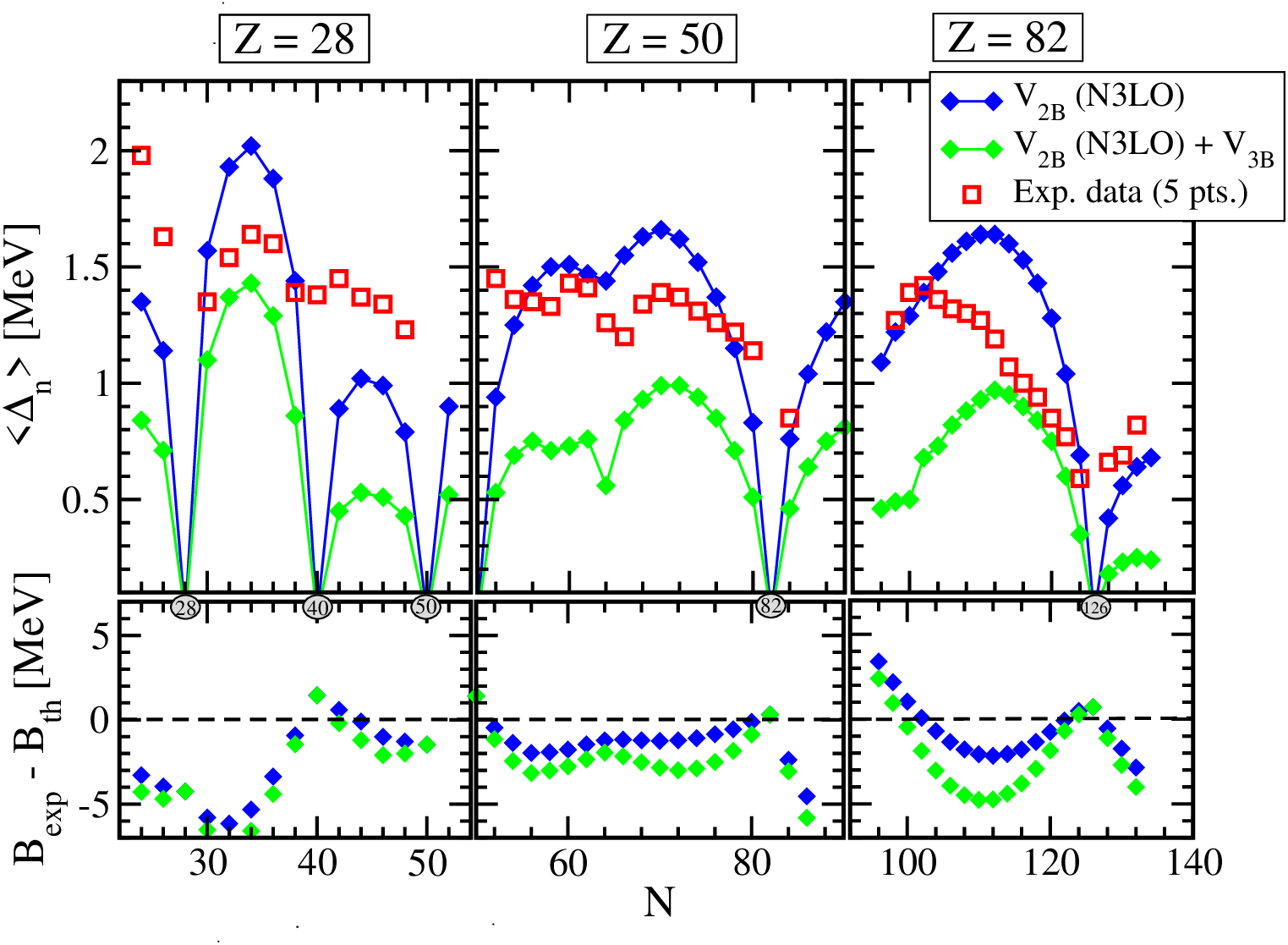}
\caption{\label{fig2}
 Theoretical average neutron pairing gaps (\ref{av-gap}) of the 
even-even isotopes of nickel $Z=28$, tin $Z=50$, and lead $Z=82$, compared to the 
empirical values calculated from experimental masses using 
Eq.~(\ref{delta5}) 
(upper panel). Absolute deviations of the calculated binding energies from the 
experimental values ~\cite{AW.03} (lower panel). 
}
\end{center}
\end{figure}

\begin{figure}[h]
\begin{center}
\includegraphics[scale=0.6,angle=0.0]{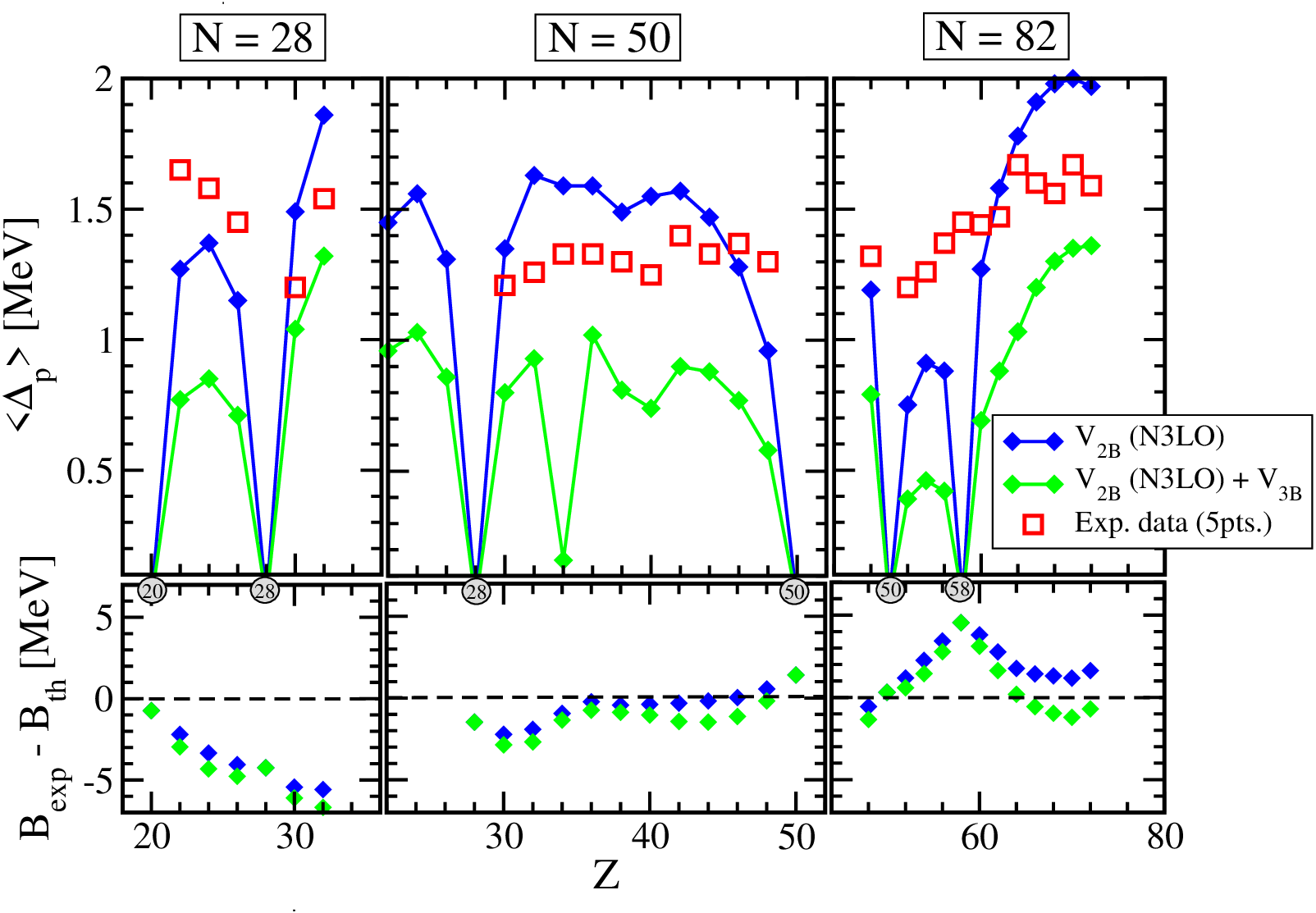}
\caption{\label{fig3}
Same as described in the caption to Fig.~\ref{fig3} 
but for the chains of 
isotones $N=28$, $N=50$ and $N=82$.}
\end{center}
\end{figure}

Employing the RHB model with the FKVW functional in the $ph$ channel 
and  the separable pairing force Eq.~(\ref{pp-force}) in the $pp$ channel, we have calculated 
the self-consistent ground-state solutions for several sequences of isotopes (nickel, 
tin and lead) and isotones ($N=28$, $N=50$ and $N=82$). The total 
binding energies and average pairing gaps are compared to available data in 
Figs.~\ref{fig2} and \ref{fig3}. The experimental masses are from  
Ref.~\cite{AW.03}, and the average proton and neutron gaps \cite{Bender:2000xk} 
\begin{equation}
\bar \Delta = \frac{ \sum_k \Delta_k u_k v_k}{\sum_k v^2_k}
\label{av-gap}
\end{equation}
are compared to empirical values determined using the 5-point formula \cite{Moller:1992zz} for even-even nuclei
\begin{equation}
\label{delta5}
\Delta^{(5)} (N_0) = -\frac{1}{8} \left[ E(N_0+2) - 4E(N_0+1) + 6E(N_0) - 4E(N_0-1) + E(N_0-2) \right] \;.
\end{equation}
$E(N_0)$ denotes the experimental binding energy of a nucleus with 
$N_0$ neutrons ($Z_0$ for protons). In Eq.~(\ref{av-gap}) the sum is over 
proton or neutron canonical states, $\Delta_k$ is the diagonal matrix element 
of the pairing field in the canonical state $k$, and $v_k$ denotes the corresponding eigenvalue of the one-body density matrix (occupation factor)\footnote{
In Refs. \cite{Heb.09,Pair3b,Lesinski:2011rn} a somewhat different prescription was used for the pairing gaps.  
The theoretical gap $\Delta_{LCS}$ (Lowest Canonical State) corresponds to the diagonal pairing matrix element $\Delta_i$ 
in the canonical single-particle state $\phi_i$ whose quasi-particle energy is the lowest, whereas
experimental gaps are deduced from binding energies using three-point mass differences centered on odd-mass nuclei.
}. 

The theoretical gaps shown in Figs.~\ref{fig2} and \ref{fig3} 
have been calculated using the values of the parameters 
 $G$ and $a$ that correspond to the nuclear matter pairing gaps in Fig.~\ref{fig1}. 
The gaps calculated by including only the interaction $V_{2B}$ (blue diamonds)
reproduce on a quantitative level the isotopic and isotopic trends of the empirical gaps. 
Including the three-nucleon interaction $V_{3B}$ induces a sizable reduction of the calculated gaps (green diamonds), 
and we note that similar conclusions were also reached in Ref. \cite{Lesinski:2011rn}. 
The calculated gaps for the isotopic chains $Z=28$, $Z=50$ and $Z=82$ indicate that 
missing contributions like, for instance particle-vibration coupling, could play an important 
role \cite{Barranco:2005yk,Gori:2005ym}. Fig.~\ref{fig3} displays similar 
results for the proton pairing gaps of the isotonic chains $N=28$, $N=50$ and $N=82$ 
(we note that here the contribution of the Coulomb interaction in the pairing channel is neglected).
The subshell closures that appear at $N=40$ in the nickel chain \cite{Broda}, and
at $Z=58$ in the $N=82$ chain \cite{Long:2006dj}, lead to a strong reduction of pairing correlations 
in the corresponding calculated ground-states. 

The influence of three-body forces on the total binding energy is much less pronounced, 
as shown in the lower panels of Figs.~\ref{fig2} and \ref{fig3}, where we display 
the absolute deviations of the calculated binding energies from the experimental values. 
On the one hand this is because pairing correlations contribute much less than the 
mean-field self-energies to the total binding. On the other hand this is a well known 
characteristic of a self-consistent calculation in that, for a given nucleus, a reduction 
of pairing results in an effective enhancement of the mean-field contribution 
to the total energy, and vice versa.  In general, the combination of the FKVW $ph$ 
effective interaction and the separable pairing force Eq.~(\ref{pp-force}) produces 
results for the total binding energies that are comparable to those obtained with the best 
empirical non-relativistic and relativistic energy density functionals. For the nickel isotopes 
the largest deviations are in the region $Z \approx N$, where one expects additional 
contributions from proton-neutron correlations that are not explicitly included in the FKVW 
functional. In the tin isotopes the calculated masses start deviating from data in 
neutron-rich nuclei beyond the major shell closure at $N=82$, whereas for lead nuclei 
the deviations are most pronounced in the lightest, neutron-deficient isotopes that are 
characterized by soft potentials and shape coexistence.  

\section{Conclusions}
\label{sec_con}

A consistent microscopic approach to the structure of open-shell  
nuclei has been introduced, in which both the {\it ph} and the {\it pp} channels of the effective 
nuclear interaction are fully determined by chiral pion-nucleon dynamics.
By employing an ansatz for the pairing force that is separable in momentum space, 
we have performed an efficient mapping of the chiral potential in 
the pairing channel (at the N$^3$LO order and the N$^2$LO in the two-body and 
three-body sectors, respectively)  to an effective $pp$ interaction for finite nuclei. 
The two parameters of the separable pairing force are adjusted to reproduce the density 
dependence of the pairing gaps in symmetric nuclear matter. 
The resulting effective pairing interaction thus enables, 
on the one hand, the treatment of pairing correlations in finite nuclei using 
pairing functionals constrained by chiral dynamics and, on the other hand, calculations 
in the $pp$ channel with a finite-range interaction. The significant advantage is that 
the computational cost is greatly reduced when compared to nonlocal finite-range forces like, 
for instance, the empirical Gogny force. A noteworthy result of the present 
investigation is that it confirms the important role of three-body forces in  
determining pairing gaps in finite nuclei.

This work was partly supported by 
INFN, MIUR and MZOS (project 1191005-1010). We acknowledge useful discussions with
N. Kaiser, W. Weise and E. Vigezzi, and would like to thank J. W. Holt for providing numerical
values for the three-body potential.

\newpage

\newpage

\begin{thebibliography}{00}

\bibitem{Bender:2003jk}
  M.~Bender, P.~H.~Heenen and P.~G.~Reinhard,
  Rev.\ Mod.\ Phys.\  {\bf 75} (2003) 121.
  
\bibitem{review}
{\it Focus Issue on Open Problems in Nuclear Structure Theory}, J. Phys. G
{\bf 37} (2010) Vol. 6.

\bibitem{DG.90} R. M. Dreizler and E. K. U. Gross,  \textit{Density Functional Theory}, Springer, Berlin (1990); E. Engel and R. M. Dreizler, {\it Density Functional Theory: An Advanced Course},
Springer, Berlin (2011).

\bibitem {FKV.03} P. Finelli, N. Kaiser, D. Vretenar and W. Weise, Eur. Phys.
J. A {\bf 17} (2003) 573.

\bibitem {FKV.04} P. Finelli, N. Kaiser, D. Vretenar and W. Weise, Nucl. Phys.
A {\bf 735} (2004) 449.

\bibitem {FKV.06} P. Finelli, N. Kaiser, D. Vretenar and W. Weise, Nucl. Phys.
A {\bf 770} (2006) 1.

\bibitem {Fri.04}S. Fritsch, N. Kaiser and W. Weise, Nucl. Phys. A {\bf 750} (2005) 259.

\bibitem {VALR.05} D. Vretenar, A.~V. Afanasjev, G.~A. Lalazissis and P. Ring,
	Phys. Rep. {\bf 409} (2005) 101.

\bibitem {BGG.91} J.~F. Berger, M. Girod and D. Gogny, Comput. Phys. Commun. {\bf 63} (1991) 365.

\bibitem{Duguet:2007be}
  T.~Duguet and T.~Lesinski, Eur.\ Phys.\ J.\ ST {\bf 156} (2008) 207.

\bibitem{Les.09} T. Lesinski, T. Duguet, K. Bennaceur and J. Meyer, 
		Eur. Phys . J. A {\bf 40} (2009) 121.
		
\bibitem{Heb.09} K. Hebeler, T. Duguet, T. Lesinski and A. Schwenk,
		 Phys. Rev. C {\bf 80} (2009) 044321. 

\bibitem{Pair3b}  T. Duguet, T. Lesinski, K. Hebeler and A. Schwenk, 
Mod. Phys. Lett. A {\bf 25} (2010) 1989.

\bibitem{Lesinski:2011rn}
  T.~Lesinski, K.~Hebeler, T.~Duguet and A.~Schwenk,
  arXiv:1104.2955 [nucl-th].

\bibitem{KNV.05} N. Kaiser, T. Nik{\v s}i{\' c} and D. Vretenar, 
			Eur. Phys . J. A {\bf 25} (2005) 257.

\bibitem{Machleidt:2011zz}
  R.~Machleidt and D.~R.~Entem,
  Phys.\ Rept.\  {\bf 503 } (2011)  1.
  
\bibitem{Epelbaum:2008ga}
  E.~Epelbaum, H.~-W.~Hammer and U.~-G.~Meissner,
  Rev.\ Mod.\ Phys.\  {\bf 81 } (2009)  1773.  
  
  \bibitem{Holt:2009ty}
  J.~W.~Holt, N.~Kaiser and W.~Weise,
  Phys.\ Rev.\  C {\bf 81 } (2010)  024002.
  
  \bibitem{Hebeler:2009iv}
  K.~Hebeler and A.~Schwenk,
  Phys.\ Rev.\  C {\bf 82 } (2010)  014314.
  
  \bibitem{Hebeler:2010xb}
  K.~Hebeler, S.~K.~Bogner, R.~J.~Furnstahl, A.~Nogga and A.~Schwenk,
  Phys.\ Rev.\  C {\bf 83 } (2011)  031301.
  
  \bibitem{Gandolfi}
  S. Gandolfi, A. Yu. Illarionov, F. Pederiva, K. E. Schmidt and S. Fantoni, 
  Phys.\ Rev.\  C {\bf 79 } (2009)  054005.
  
  \bibitem{Lovato:2010ef}
  A.~Lovato, O.~Benhar, S.~Fantoni, A.~Y.~Illarionov and K.~E.~Schmidt,
  Phys.\ Rev.\  C {\bf 83} (2011) 054003
  
\bibitem {TMR.09a} Y. Tian, Z.~Y. Ma and P. Ring, Phys. Lett. B {\bf 676} (2009) 44.

\bibitem {TMR.09b} Y. Tian, Z.~Y. Ma and P. Ring, Phys. Rev. C {\bf 79} (2009) 064301.

\bibitem{Mig.67} A.B. Migdal, {\it Theory of finite Fermi systems and applications to atomic nuclei}, Interscience, New York (1967).    

\bibitem{DH-J.03} D.J. Dean and M. Hjorth-Jensen, Rev. Mod. Phys. {\bf
75}  (2003) 607; and references therein.

\bibitem{Sch.96} H.J. Schulze, J. Cugnon, A. Lejeune, M. Baldo and U. Lombardo,
                Phys. Lett. B {\bf 375}, (1996) 1.
                
\bibitem{Lombardo:2005sw}
  U.~Lombardo, H.~Schulze, C.~-W.~Shen and W.~Zuo,
  Int.\ J.\ Mod.\ Phys.\  E {\bf 14 } (2005)  513.
  
\bibitem{Holt:2011nj}
  J.~W.~Holt, N.~Kaiser and W.~Weise,
  arXiv:1107.5966 [nucl-th].  
  
  \bibitem{Fritsch:2004nx}
  S.~Fritsch, N.~Kaiser and W.~Weise,
  Nucl.\ Phys.\  A {\bf 750 } (2005)  259.

\bibitem{Holt:2011nj2}
  J.~W.~Holt, N.~Kaiser and W.~Weise,
  arXiv:1106.5702 [nucl-th].  

\bibitem{Sedrakian:2003cc}
  A.~Sedrakian, T.~T.~S.~Kuo, H.~Muther and P.~Schuck,
  Phys.\ Lett.\  B {\bf 576} (2003)  68.

\bibitem{AW.03} G. Audi, A. H. Wapstra and C. Thibault,
	Nucl. Phys. A {\bf 729} (2003) 337.

\bibitem{Bender:2000xk}
  M.~Bender, K.~Rutz, P.~G.~Reinhard and J.~A.~Maruhn,
  Eur.\ Phys.\ J.\  A {\bf 8} (2000) 59.

\bibitem{Moller:1992zz}
  P.~Moller and J.~R.~Nix,
  Nucl.\ Phys.\  A {\bf 536} (1992) 20.

\bibitem{Barranco:2005yk}
  F.~Barranco, P.~F.~Bortignon, R.~A.~Broglia, G.~Colo, P.~Schuck, E.~Vigezzi and X.~Vinas,
  Phys.\ Rev.\ C {\bf 72 } (2005)  054314.

\bibitem{Gori:2005ym}
  G.~Gori, F.~Ramponi, F.~Barranco, P.~F.~Bortignon, R.~A.~Broglia, G.~Colo' and E.~Vigezzi,
  Phys.\ Rev.\ C {\bf 72 } (2005)  011302.

\bibitem{Broda}
R.~Broda, B.~Fornal, W.~Krolas, T.~Pawlat, D.~Bazzacco, S.~Lunardi, C.~Rossi-Alvarez, R.~Menegazzo {\it et al.},
  Phys.\ Rev.\ Lett.\  {\bf 74 } (1995)  868. 
 
\bibitem{Long:2006dj}
  W.~H.~Long, H.~Sagawa, J.~Meng and N.~Van Giai,
  Europhys.\ Lett.\  {\bf 82 } (2008) 12001.
 
\end{thebibliography}
\end{document}